\documentclass[12pt]{article}
\raggedright
\newcommand{\T}{{\rm tr}}

\newcommand{\cH}{{\cal H}}

\newcommand{\1}{{\bf 1}}
\newcommand{\rc}{{\rm anti}}
\begin{document}
\title{ On 1--qubit channels }

\author{Armin Uhlmann }

\date{Institut f.~Theoretische Physik, Universit\"at Leipzig\\
Augustusplatz 10/11, D-04109 Leipzig}

\maketitle

\begin{abstract}
The entropy $H_T(\rho)$ of a state with respect to a channel
$T$ and the  Holevo capacity of the channel
require the solution of difficult variational
problems. For a class of 1-qubit channels, which contains
all the extremal ones, the problem can be significantly
simplified by associating an Hermitian anti-linear operator
$\vartheta$ to every channel of the considered class. The channel's
concurrence $C_T$ can be expressed by $\vartheta$ and
turns out to be a flat roof. This allows to write down
an explicit expression for $H_T$. Its maximum would give
the Holevo (1--shot) capacity.

\end{abstract}



\section{Introduction}
Given two Hilbert spaces, $\cH^{\rm in}$ and $\cH^{\rm out}$,
of finite dimension,
a quantum channel is a completely positive and trace preserving
linear map,
\begin{equation} \label{d1}
T \, : \, \, \rho^{\rm in} \longrightarrow
\rho^{\rm out} := T(\rho^{\rm in})
\end{equation}
from the operators of $\cH^{\rm in}$ into those
of $\cH^{\rm out}$.
To shorten notations, we skip the in and out superscripts
mostly. The {\em rank} of $T$ is the maximal rank within
all output density operators $T(\rho)$.

An important device to estimate, how effectively a channel works,
is the quantity
\begin{equation} \label{d2}
H_T(\rho) =  \max \sum p_j S[ T(\rho) \parallel T(\rho_j) ]
\end{equation}
In this variational problem one has to compare all
convex decompositions
\begin{equation} \label{d3}
\rho = \sum p_j \rho_j, \quad p_j \geq 0
\end{equation}
of the input density operator $\rho$ into other input density
operators $\rho_j$, and $S(. \parallel .)$ abbreviates the
relative entropy. According to Holevo \cite{Ho73},
the quantity (\ref{d2})
can be interpreted as the maximum of mutual information between
the input and the output of $T$ for a given ensemble average
$\rho$. In Benatti \cite{Be96} it is identified with the maximal
accessible information for all quantum sources with ensemble
average $\rho$. In Schumacher and Westmoreland
\cite{SW99}, where the problem is considered even
in a more general context, this maximum is denoted by $\chi^*$,
and an ensemble saturating (\ref{d2}) is called an {\em optimal
signal ensemble.}
One gets the {\em Holevo} or {\em one shot capacity} by
$$
{\bf C}(T) = \max_{\rho} H_T(\rho)
$$
On the other hand, (\ref{d2}) is a decisive tool for the construction
of the CNT--entropy of Connes, Narnhofer and Thirring \cite{CNT87}.
In this context it is called
{\em entropy of $\rho^{\rm in}$ with respect to the channel
$T$}, an appropriate generalization of the
{\em entropy of $\rho$ with respect to subalgebra},
\cite{NT85}.
(I apologize for the change in notation:
In \cite{CNT87}
and \cite{Uh98} the quantity $H_T(\rho)$ has been called
$H_{\rho}(T)$.)
By the von Neumann entropy
$$
S_T(\rho) = S[ T(\rho) ]
$$
of $T(\rho)$, (\ref{d2}) can be rewritten as
$$
H_T(\rho) = S_T(\rho) - \min \sum p_j S_T(\rho_j)
$$
The concavity of $S$ allows to restrict the convex decompositions
(\ref{d3}) to the extremal ones, i.~e. to those consisting of
pure input states only. This simple observation implies that
we may write
\begin{equation} \label{d4}
H_T(\rho) = S_T(\rho) - E_T(\rho)
\end{equation}
where
\begin{equation} \label{d5}
E_T(\rho) = \min \sum p_j S_T(\pi_j), \quad \pi_j \, \hbox{pure}
\end{equation}
under the condition $\rho = \sum p_j \pi_j$.
The simplification
is not only by the restriction of the original variation
to decompositions with pure states. $E_T$
enjoys in addition the {\em roof property:}
If one can find an optimal decomposition of $\rho$ into
pure states saturating (\ref{d5}), then $E_T$ is
convexly linear on the convex subset generated by the
pure input states $\pi_j$ (provided $p_j > 0$).
Moreover, $E_T$ is convexly linear on
the convex set generated by the union of
{\em all} those pure input states $\pi$ which
can occur in any optimal decomposition of a given input
density operator $\rho$,  see \cite{Uh98} or \cite{Uh00c} for
more details.

For the 1--qubit channel, which substitutes
the off--diagonal elements of $\rho$ by zeros,
the first published computation of $E_T$ and $H_T$
I know is by Levitin, \cite{Le94}. He explicitly points to
the constancy of $E_T$ along the straight lines
of density operators parallel to the 3--axis with
fixed off--diagonal elements.
But before knowing Levitin's work I have seen this
surprising feature by a computer program of R.~F.~Werner
(Vienna 1994).
An instructive example, how this
observation can be used to compute parts of $E_T$ of
the same problem but for rank three, is in
Benatti et all, \cite{BNU96} and \cite{BNUx}.

On the
other hand, if $T$ is the relative trace to the states
of one part in a bipartite system, $E_T$ is
the {\em entanglement of formation} of
Bennett et all \cite{BDSW96}. These authors compute $E_T$
for Werner states \cite{We89}
in 2-by-2 dimensions. Higher dimensional
Werner states are treated in
Terhal and Vollbrecht \cite{TV00} and Vollbrecht and Werner
\cite{VW00}.
As a heuristic  guide, one may think of $E_T(\rho)$
a measure of entanglement of $\rho$ with respect to
an arbitrary channel $T$.

The convex roof extension,
$E_T$, is the largest convex function on the input density
operators coinciding with $S_T$ for pure
states $\pi = \pi^{\rm in}$. As $-E_T$ is concave,
the computation of $H_T$ can be paraphrased as following:

{\em Add to $S_T$ the smallest concave function of the input
states such that the sum vanishes at pure input states.
One obtains $H_T$.}

Now I pass to a quite different topic. In certain cases
one can effectively compute channel characteristics by
anti-linear operators.
An anti-linear operator $\vartheta$ acts on kets
according to
$$
\vartheta \sum a_j |j \rangle = \sum
a_j^* \vartheta |j \rangle
$$
With respect to a basis, $\vartheta$ is completely described by its
matrix elements $\langle j| \vartheta |k \rangle$ and, to
distinguish its matrix representation from the linear
situation, I add an index {\em anti} to it. Hence, in two
dimension, let us write
$$
\vartheta = \pmatrix{\alpha_{00} & \alpha_{01} \cr
\alpha_{10} & \alpha_{11} }_{\rc}
$$
for the matrix representation based on $|0\rangle$ and $|1\rangle$.
As a merit we easily can compute matrix products. For instance
$$
\{ \alpha_{jk} \}_{\rc} \cdot \{ a_{li} \} =
\{ \sum_k \alpha_{jk} a_{ki}^* \}_{\rc}
$$
represents the product of an anti-linear and a linear operator,
which is again anti-linear. The hermitian adjoint, $\vartheta^{\dag}$,
of an anti-linear $\vartheta$ is again anti-linear and defined
by the rule
$$
\langle j| \vartheta^{\dag} |k \rangle =
\langle k| \vartheta |j \rangle
$$
The Hermitian conjugate thus changed the matrix elements
of an anti-linear operator from $\alpha_{jk}$ to $\alpha_{kj}$.
It follows that $\vartheta \to \vartheta^{\dag}$ is a
{\em linear} operation for anti-linear operators, quite in
contrast to the linear case.
We shall mostly need Hermitian
(i.e. self-adjoint) anti-linear operators. The matrix entries
for these operators are characterized by the symmetry
condition $\alpha_{jk} = \alpha_{kj}$, and by nothing else.

There are some warnings concerning the use of anti-linear
operators and maps. Two of them are: 1) Do not apply them to bras,
i.e. from right to left: An expression like
$\langle y| = \langle x| \vartheta$ is ill--defined.
2) One cannot tensor an anti-linear operator with
a linear one.

Next,
$\vartheta \vartheta^{\dag}$ and $\vartheta^{\dag} \vartheta$
are positive linear operators with equal eigenvalues.
The eigenvalues of an anti-linear operator itself, however,
fill some circles, see Wigner \cite{Wi60} for more details.
Therefore, the determinant is defined only up to a phase factor.
The trace is undefined for anti-linear operators.

\section{Rank two channels}
Let us now assume $\dim \cH^{\rm out} = 2$. Then there is only one
free variable on which $S_T(\rho)$ depends. This fact has been
used already in \cite{BDSW96} and, with a very remarkable
result, in Hill and Wootters \cite{HilWoo97}
and Wootters \cite{Woo97}, to compute
the entanglement of formation in the  2--qubit case.

Following \cite{BDSW96}, the first issue
is to start with a suitable expression for $S_T$. With
$$
h(x) = - x \ln_2 x - (1-x) \ln_2 (1-x)
$$
one defines
\begin{equation} \label{f1}
f(y) := h({1 + \sqrt{1 - y^2} \over 2})
\end{equation}
which is increasingly monotone
in $0 \leq y \leq 1$ from $0$ to $1$, and
convex in $-1 \leq y \leq 1$. (One checks that the
first derivative, $f'$, is increasing.)
For a 2-by-2 density operator $\omega$ with eigenvalues
$\mu_1 \geq \mu_2$ one gets
\begin{equation} \label{f2}
S(\omega) = h(\mu_1) = f(y), \quad y = 2 \sqrt{\det \omega}
\end{equation}
which can be seen from
$$
1 - y^2 = (\mu_1 + \mu_2)^2 - 4 \mu_1 \mu_2
= (\mu_1 - \mu_2)^2
$$
It follows
\begin{equation} \label{f2a}
S_T(\rho) = f[ 2 \sqrt{ \det T(\rho)}]
\end{equation}

Our next task is to define the convex roof $C_T$ to be
the largest convex function on $\cH^{\rm in}$ which
coincides for pure input states $\pi$ with
$\sqrt{\det T(\pi)}$. The letter $C$ and
the name {\em concurrence of $T$} for $C_T$
I borrowed from \cite{BDSW96} and from \cite{Woo97}.
To give an equation,
\begin{equation} \label{f3}
C_T(\rho) =  \min \sum p_j \sqrt{ \det T(\pi_j) }
\end{equation}
the minimum is running through all convexly linear
decompositions $\sum p_j \pi_j$ of $\rho$ with
pure input states. As a matter of fact, one
cannot beat this minimum in allowing the $\pi_j$
to become mixed. This is due to the concavity of
$\sqrt{\det \omega}$ in two dimensions.
(In the language of convex analysis: The convex hull
of a concave function is a roof, see \cite{Uh98},
appendix.) As a by--product
$$
C_T(\rho) \leq \sqrt{ \det T(\rho) }
$$
The range of $C_T$ is from 0 to 0.5, and it is convex by
definition. Because $f$ in (\ref{f1}) is convex and
increasing, the function
\begin{equation} \label{f3a}
\rho \longrightarrow f[2 C_T(\rho) ]
\end{equation}
is a convex function which equals $S_T$ for pure states.
Though $C_T$ is a roof, this is not sufficient for proving
the equality of (\ref{f3a}) with $E_T$. Why should a function
of a roof remain a roof? There is no general reason for that.
There exists, however, one special case not burden with the
mentioned difficulty: Let us call $C_T$ {\em flat} if there
is, for every $\rho$,  an optimal pure state decomposition
$$
\rho = \sum p_j \pi_j, \quad C_T(\rho) = \sum
p_j C_T(\pi_j)
$$
such that
\begin{equation} \label{f4}
C_T(\pi_1) = C_T(\pi_2) = \dots = C_T(\pi_j) = \dots
\end{equation}
If this takes place, every $\rho$ is contained in a convex subset
which is generated by pure input states, and on which
the roof is not only linear but even constant.

Thus, if we would know the flatness of $C_T$, every function
of it must be a roof, though not necessarily a convex one.
But the convexity of (\ref{f3a}) has been stated already.
Altogether one arrives at

{\bf Lemma 1:} \, If the roof $C_T$ is flat then
\begin{equation} \label{f5}
E_T(\rho) = f[ 2 C_T(\rho) ]
\end{equation}

We are faced with two problems: How to compute $C_T$, and
how to check whether it is a flat roof.
The next aim is to give a large class of rank two channels
fulfilling the desired flatness condition.

\section{1--qubit channels of length two}
Let $\cH$ be of dimension two, and $T$ a quantum channel
of the form
\begin{equation} \label{k1}
T(\rho) = A \rho A^{\dag} + B \rho B^{\dag}
\end{equation}
The set of channels mapping the 1--qubit density operators
into themselves is convex. Its structure is well described in
King and Ruskai \cite{KR99} and in
Ruskai et al. \cite{RSW00},  where a complete list of all its
extremal maps has been given. As shown in \cite{RSW00},
every extremal 1--qubit channel has a representation (\ref{k1}).
We may, for example, choose
\begin{equation} \label{k2}
A = \pmatrix{a_{00} & 0 \cr 0 & a_{11}}, \quad
B = \pmatrix{0 & b_{01} \cr b_{10} & 0}
\end{equation}
To be trace preserving one has to have
$$
|a_{00}|^2 + |b_{10}|^2 = |a_{11}|^2 + |b_{01}|^2 = 1
$$
According to \cite{RSW00}, one can choose $A$ and $B$
in (\ref{k2}) with real entries to get all the
extremal maps up to unitary equivalence. We are going to
prove:

{\em For all quantum channels of the form (\ref{k1}) $C_T$
is flat, and there exist explicit expressions for $C_T$, $E_T$,
and $H_T$.}  One of the two key observations is

{\bf Theorem 2:} \,  Given a super-operator as in (\ref{k1}).
There is an Hermitian anti-linear operator, $\vartheta$,
such that
\begin{equation} \label{2ch2}
\det T(\pi) = \T \, \pi (\vartheta \pi \vartheta)
\end{equation}
is true for all pure density operators $\pi$.

{\bf Proof.} The proof of the theorem goes in three steps.
In the first two, both sides of (\ref{2ch2}) are computed.
The last one is a comparison of the results.

Let be $a_{jk}$, $j,k = 0,1$,
the matrix elements of $A$ wit respect of a reference basis.
Accordingly let us write $B = \{ b_{jk}\}$. The application
of $A$ and $B$ to a vector $\{ x_0, x_1\}$  is called
$\{z_0, z_1\}$ and $\{w_0, w_1\}$ respectively. Hence
$$
T( \pmatrix{x_0 x_0^* & x_0 x_1^* \cr x_1 x_0^* & x_1 x_1^*} )
= \pmatrix{z_0 z_0^* + w_0 w_0^* & z_0 z_1^* + w_0 w_1^* \cr
z_1 z_0^* + w_1 w_0^* & z_1 z_1^* + w_1 w_1^* }
$$
The determinant is given by
\begin{equation} \label{th1.1}
\det
T( \pmatrix{x_0 x_0^* & x_0 x_1^* \cr x_1 x_0^* & x_1 x_1^*} )
=
(z_0 w_1 - z_1 w_0) (z_0 w_1 - z_1 w_0)^*
\end{equation}
From
$z_0 = a_{00} x_0 + a_{01} x_1$, $w_1 = b_{10} x_0 + b_{11} x_1$,
and so on,
we get the $w_j$ by using the coefficients $b_{jk}$.
Hence
\begin{equation} \label{th1.2}
z_0 w_1 - z_1 w_0 = c_{00} x_0^2 + c_{11} x_1^2 +
(c_{01} + c_{10}) x_0 x_1
\end{equation}
where
$$
c_{00} = a_{00} b_{10} - a_{10} b_{00}, \quad
c_{11} = a_{01} b_{11} - a_{11} b_{01}
$$
\begin{equation} \label{th1.3}
c_{01} + c_{10} = a_{00} b_{11} + a_{01} b_{10} - a_{10} b_{01}
- a_{11} b_{00}
\end{equation}
Let us now consider step two of the proof. An
anti-linear operator, $\vartheta$, can be characterized
by the entries of its matrix representation in a
given reference basis. Let us denote $\vartheta$ and its
Hermitian adjoint by
\begin{equation} \label{th1.4}
\vartheta = \pmatrix{ \alpha & \beta \cr \gamma & \delta}_{\rc},
\quad \vartheta^{\dag} =
\pmatrix{ \alpha & \gamma \cr \beta & \delta}_{\rc}.
\end{equation}
For a general density operator, $\rho$, with entries
$\rho_{jk}$ in the reference basis, one obtains
$$
\vartheta^{\dag} \rho \vartheta = \pmatrix{
\rho_{00} \alpha \alpha^* + \rho_{10} \alpha \gamma^* +
\rho_{01}  \gamma \alpha^* + \rho_{11} \gamma \gamma^* &
\rho_{00} \alpha \beta^*  + \rho_{10} \alpha \delta^* +
\rho_{01} \gamma \beta^*  + \rho_{11} \gamma \delta^* \cr
\rho_{00} \beta \alpha^*  + \rho_{10} \beta \gamma^* +
\rho_{01} \delta \alpha^* + \rho_{11} \delta \gamma^* &
\rho_{00} \beta \beta^*   + \rho_{10} \beta \delta^* +
\rho_{01} \delta \beta^*  + \rho_{11} \delta \delta^*
}
$$
It follows
$$
\vartheta^{\dag}
\pmatrix{x_0 x_0^* & x_0 x_1^* \cr x_1 x_0^* & x_1 x_1^*}
\vartheta=
\pmatrix{
(\alpha x_0^* + \gamma x_1^*) (\alpha^* x_0 + \gamma^* x_1) &
(\alpha x_0^* + \gamma x_1^*) (\beta^* x_0 + \delta^* x_1)  \cr
(\beta x_0^* + \delta x_1^*)  (\alpha^* x_0 + \gamma^* x_1) &
(\beta x_0^* + \delta x_1^*) (\beta^* x_0 + \delta^* x_1)
}
$$
and, finally,
\begin{equation} \label{th1.6}
\T \,
\pmatrix{x_0 x_0^* & x_0 x_1^* \cr x_1 x_0^* & x_1 x_1^*}
\vartheta^{\dag}
\pmatrix{x_0 x_0^* & x_0 x_1^* \cr x_1 x_0^* & x_1 x_1^*}
\vartheta =
| x_0 (\alpha^* x_0 + \gamma^* x_1) +
x_1 (\beta^* x_0 + \delta^* x_1) |^2
\end{equation}
Comparing with (\ref{th1.3}) the determinant of $T(\pi)$ is
equal to the trace (\ref{th1.6}) if
\begin{equation} \label{th1.7}
\alpha^* = c_{00}, \, \, \beta^* + \gamma^* = c_{01} + c_{10},
\, \, \delta^* = c_{11}
\end{equation}
With this choice we have
\begin{equation} \label{th1.8}
z_0 w_1 - w_1 z_0 = \langle \phi | \vartheta | \phi \rangle^*
\end{equation}
Now we impose hermiticity. $\vartheta$ is Hermitian if and only if
$\beta = \gamma$. We see from (\ref{th1.7}) that there is exactly
one Hermitian anti-linear $\vartheta$ with which (\ref{2ch2}) is
satisfied. This proves the theorem.

Before going ahead, let us write down $\vartheta$ for
the subset of channels with Kraus operators (\ref{k2}).
Denoting the matrix entries as in (\ref{th1.4}) we get
$\beta = \gamma = 0$ and
\begin{equation} \label{k2.1}
\alpha = a_{00}^* b_{10}^*, \quad \delta = - a_{11}^* b_{01}^*
\end{equation}

To get the last piece of the puzzle I recall, as an adoption
of \cite{Woo97}, a definition of \cite{Uh00c}. Define,
for two general density operators $\omega_1$ and $\omega_2$,
\begin{equation} \label{co1}
C(\omega_1, \omega_2) := \max \{ 0, \, \lambda_1 -
\sum_{j > 1} \lambda_j \}
\end{equation}
where the lambdas are the decreasingly ordered eigenvalues
of
$$
  \bigl( \sqrt{\omega_1} \omega_2 \sqrt{\omega_1} \bigr)^{1/2}
$$
If $\omega_1$ and $\omega_2$ are both of rank two, there
are not more than two non--zero eigenvalues. This reduces
(\ref{co1}) to $|\lambda_1 - \lambda_2|$, and one
obtains, \cite{Uh00c}, the expression
\begin{equation} \label{co2}
C(\omega_1, \omega_2)^2 = \T \, \omega_1 \omega_2 - 2
\sqrt{\det \omega_1 \det \omega_2}
\end{equation}
There is a general feature of (\ref{co1}),
so to say the door for the key given by theorem 2, and
which is proved in \cite{Uh00c}:

{\bf Theorem 3:} \, Let $\vartheta$ be an anti-linear Hermitian
operator in an Hilbert space. The function
\begin{equation} \label{co3}
\omega \longrightarrow C(\omega, \vartheta \omega \vartheta)
\end{equation}
is a flat convex roof on the set of density operators.

Now, returning to our 1-qubit channels, let us look for
the values of (\ref{co3}) for a pure state
$\pi = |\phi\rangle\langle\phi|$. By (\ref{co2}) it is really
easy to see that
\begin{equation} \label{co4}
C(\pi, \vartheta \pi \vartheta)^2 = \T \, \pi \vartheta \pi \vartheta
= |\langle \phi | \vartheta | \phi \rangle|^2
\end{equation}

By combining theorems 2 and 3 the structure of
$E_T$ for the channels (\ref{k1}) becomes evident. By
theorem 2 we find
$$
\det T(\pi) = C( \pi, \vartheta \pi \vartheta )^2
$$
and, finally,
\begin{equation} \label{co5}
C_T(\rho)^2 =  C(\rho, \vartheta \rho \vartheta)^2 =
 \T \, (\rho \vartheta \rho \vartheta) - 2 \det \rho \,
\sqrt{\det (\vartheta^2)},
\end{equation}
\begin{equation} \label{co6}
E_T(\rho) = f[ 2 C(\rho, \vartheta \rho \vartheta) ],
\end{equation}
and this is the solution of the variational problem we
looked for.

{\em Examples.} For the channels with Kraus operators
(\ref{k1}) the expression (\ref{co5}) can be made more
explicit. In this case the matrix representation of $\vartheta$
is diagonal with entries (\ref{k2.1}). Hence
$$
\vartheta \rho \vartheta = \pmatrix{
\rho_{00} \alpha \alpha^* &
\rho_{10} \alpha \delta^* \cr
\rho_{01} \delta \alpha^* &
\rho_{11} \delta \delta^* }
$$
$$
\T (\rho \vartheta \rho \vartheta) =
\rho_{00}^{2} \alpha \alpha^* +
\rho_{10}^2 \alpha \delta^* +
\rho_{01}^2 \delta \alpha^* +
\rho_{11}^2 \delta \delta^*
$$
This we have to insert in (\ref{co5}), reminding that we have
to subtract $\det \rho$ multiplied with twice the absolute value
$|\alpha \delta|$ of $\alpha \delta$. We take a root of
$\alpha \delta^*$ and choose its complex conjugate as the
root of $\alpha^* \delta$. With this convention the following
is unambiguous.
\begin{equation} \label{k2.2}
C_T^2 = ( |\alpha| \rho_{00} - |\delta| \rho_{11} )^2 +
 ( \sqrt{\alpha \delta^* } \rho_{01} +
 \sqrt{\alpha^* \delta } \rho_{10} )^2
\end{equation}
At first let us treat the degenerate case with
\begin{equation} \label{k2.3}
A = \pmatrix{ 1 & 0 \cr 0 & \sqrt{t} }, \quad
B = \pmatrix{0 & \sqrt{1-t} \cr 0 & 0}
\end{equation}
and $1 \geq t > 0$.
Then (\ref{k2.2}) reduces to
\begin{equation} \label{k2.4}
C_T(\rho) = \sqrt{t(1-t)} \, \rho_{11}
\end{equation}
The foliation of the set of density operators induced by $C_T$ and $E_T$
is given by the intersections of the Bloch ball with the
planes perpendicular to the 3--axis. $S_T$ is the von
Neumann entropy of
\begin{equation} \label{k2.5}
T(\rho) = \pmatrix{ 1 - t \rho_{11} & \sqrt{t} \rho_{01} \cr
\sqrt{t} \rho_{10} & t \rho_{11} }
\end{equation}
The determinant of $T(\rho)$, given $\rho_{11}$,
is maximal for $\rho_{01} = 0$, and so does $S_T$. Therefore,
on a given leaf with constant $C_T$, the maximum of $S_T$
is $h(t \rho_{11})$. It follows
$$
H_T(\rho) \leq h(t \rho_{11}) - h({1 + \sqrt{1 - 4t(1-t) \rho_{11}^2}
 \over 2}) = H_T(\rho')
 $$
on the plane containing the density operators with given
$\rho_{11}$. $\rho'$ is the diagonal part of the density
operator $\rho$. Hence
\begin{equation} \label{k2.6}
{\bf C}(T) = \max_{0 \leq r \leq 1}  [ h(rt) -
h({1 - \sqrt{1 - 4t(1-t) r^2} \over 2}) ]
\end{equation}
Smolin \cite{Sm98} has shown that the maximum is not achieved
for orthogonal input states. (The first but more complicated
example is by Fuchs \cite{Fu97}.) Indeed, as long
$\rho_{11} \neq 1/2$, there are no pairs of orthogonal states
in the leaves dictated by $C_T$.

Switching to the not degenerate case, the leaves of constant
concurrence $C_T$ are the intersection of straight lines
with the Bloch ball. We get such a line by first fixing a plane
of operators with constant diagonal entries. A second plane
is obtained by constraining the off--diagonal entries to
\begin{equation} \label{k2.7}
 \sqrt{\alpha \delta^*} \rho_{01} + \sqrt{\alpha^* \delta}
\rho_{01}^* = r
\end{equation}
$r$ real. The intersection of the planes defines a line.
$C_T$ remains constant on its intersection with the
Bloch ball.

$C_T$ is zero if both terms in (\ref{k2.2}) vanish. The
line segment cuts the Bloch sphere necessarily at pure
states. That there are one or two pure states in the range
of the channels (\ref{k1}) is proved in \cite{RSW00}.

\section{A special class of 1--qubit channels}
We would like to extend the computations to some channels with
more than two Kraus operators.
It has been proved above that we can associate to every pair
of operators, interpreted as Kraus operators, an anti-linear
Hermitian one,
\begin{equation} \label{amap}
\{ \, A, \, B \, \} \, \, \longrightarrow \vartheta
\end{equation}
One may ask whether one can
change the super-operator (\ref{d1}) without changing
$\vartheta$ and, hence, without changing $C_T$ and
$E_T$. To do so, we first observe
that the trace one condition is irrelevant for theorem 2.
This fact simplifies the following a bit, and we can allow
slightly more: After changing the Kraus operators,
$\vartheta$, and hence $C_T$, may become scaled.

The answer is in the somehow surprising identity
\begin{equation} \label{product}
( A \otimes B - B \otimes A) \, | \phi \otimes \phi\rangle
= \langle \phi | \vartheta | \phi\rangle^* (|01\rangle - |10\rangle)
\end{equation}
in which
$$
| \phi \otimes \phi\rangle = x_0^2 |00\rangle + x_0 x_1
( |01\rangle + |10\rangle ) + x_1^2 |11\rangle
$$
Consequently, if the super-operator $T'$ comes with Kraus
operators $A'$ and $B'$, and if
\begin{equation} \label{product1}
A' = \mu_{11} A + \mu_{12} B, \quad B' = \mu_{21} A + \mu_{22} B
\end{equation}
then the left hand side of (\ref{product}) changes by a
factor only. The factor is the determinant of the
transformation (\ref{product1}). Remembering the definition
of $C_T$, it results
\begin{equation} \label{product2}
C_{T'} = | \mu_{11} \mu_{22} - \mu_{12} \mu_{21} | \, C_T
\end{equation}
Now let us go a step farther and consider a channel
\begin{equation} \label{k3}
T'(\rho) = \sum_{j=1}^m A_j \rho A_j^{\dag}
\end{equation}
For a small class of these channels $C_T$ and, therefore,
$E_T$ can be computed explicitly.

{\bf Theorem 4:} \, If the linear span of the Kraus
operators $A_1$, ..., $A_m$ in (\ref{k3})
is at most 2--dimensional, there is an
anti-linear and Hermitian $\vartheta'$ satisfying
\begin{equation} \label{mch2}
\det T'(\pi) = \T \, \pi (\vartheta' \pi \vartheta')
\end{equation}
for pure $\pi$, and $C_{T'}$ is a flat roof.

{\bf Proof.} We use the identity
\begin{equation} \label{identity}
\det \sum \pmatrix{a_i c_i & a_i d_i \cr b_i c_i & b_i d_i}
= \sum_{j < k} (a_i b_k - a_k b_i) (c_i d_k - c_k d_i)
\end{equation}
to compute the determinant of $Y = \det T(\pi)$,
\begin{equation} \label{ma1}
Y = \sum \pmatrix{y_{i0} y_{i0}^* &  y_{i0} y_{i1}^*
\cr  y_{i1} y_{i0}^* &  y_{i1} y_{i1}^* }
\end{equation}
where $\pi = |\phi\rangle\langle\phi|$,
$\phi = x_0 |0\rangle + x_1 |\phi\rangle$, and
\begin{equation} \label{ma2}
A_i \, \pmatrix{x_0 \cr x_1} = \pmatrix{y_{i0} \cr y_{i1}}
\end{equation}
From (\ref{identity}) we obtain
\begin{equation} \label{ma3}
\det T(\pi) = \sum_{j < k} | y_{0j} y_{1k} - y_{1j} y_{0k} |^2
\end{equation}
We choose $A$ and $B$ in (\ref{k1}) of the channel $T$
as linear generators of the linear span of the $A_j$ in
(\ref{mch2}). There are numbers $\mu_l^j$ fulfilling
\begin{equation} \label{ma4}
A_j = \mu_1^j A + \mu_2^j B, \quad j = 1, \dots, m
\end{equation}
and allowing to rewrite
$$
| y_{0j} y_{1k} - y_{1j} y_{0k} | =
| \mu_1^j \mu_2^k - \mu_2^j \mu_1^k | \cdot | z_0 w_1 - z_1 w_0 |
$$
By the help of (\ref{th1.8}) we finally obtain
\begin{equation} \label{ma5}
\det T(\pi) = ( \T \pi \vartheta \pi \vartheta ) \,
\sum_{j < k} | \mu_1^j \mu_2^k - \mu_2^j \mu_1^k |^2
\end{equation}
Hence, $\vartheta' = \mu \vartheta$, where $|\mu|^2$ can be
read off from (\ref{ma5}), does the job required
by theorem 4.

It seems, theorem 4 exhausts the possibilities to compute
$C_T$ and $E_T$ by an anti-linear and Hermitian $\vartheta$
for 1--qubit channels
in the manner of the present paper. There are simple examples
where the linear span of the Kraus operators is of
dimension larger than two and for which one cannot find
an appropriate $\vartheta$.
For instance, the well known
depolarizing channels
$$
T_t(\rho) = [ (\T \rho) \1 + s \rho ] (s + \dim \cH )^{-1}
$$
which are positive for $-1 \leq s$ and completely positive
for $- (\dim \cH)^{-1} \leq s$ belong to them.
The determinant of $T(\pi)$ is constant for pure states.
Consequently, $C_T$ is constant everywhere and, trivially,
a flat roof. But if this
constant is different from zero, i.~e. $s \neq 0$,
it cannot be
represented as (\ref{mch2}) for all pure $\pi$ even if
the dimension of $\cH$ is two.


\end{document}